# Defect-free nanocrystals are at thermodynamic equilibrium at room temperature

Faramarz Hossein-Babaei[1*] and Hamid Reza Sefidi Shirkoohi[1]

[1]Electronic Materials Laboratory, Electrical Engineering Department, K. N. Toosi University of Technology, Tehran 16315-1355, Iran

fhbabaei@kntu.ac.ir

**Abstract**

Crystal defect statistics is developed by minimizing the Gibbs free energy of defect formation, which is commonly converted to a crystallite size-independent equation by applying Stirling's approximation. Solutions of this equation forecast Arrhenius-like temperature dependence for the concentrations of all defect types, and higher defect populations for nanocrystals due to the smaller formation free energies involved. Here, we improve the accuracy in the mathematical processing of the equation describing the defect population at thermodynamic equilibrium and show that this equation is intrinsically size-dependent. The new model predicts lower defect concentrations for smaller crystallites, and shows that vacancy-free crystallites smaller than a determinable critical size are thermodynamically stable at elevated temperatures. Our findings describe the previously reported data on the mechanical properties of a number of nanocrystals and lead to a revised and deeper understanding of many morphological occurrences in micro- and nanocrystals.

*Corresponding author; fhbabaei@kntu.ac.ir

## 1. Introduction

Determining the population of point defects, particularly vacancies, in pure single crystals at thermodynamic equilibrium has been the subject of numerous theoretical and experimental studies[1-8]. The findings of these investigations are consistent in predicting that a) vacancy concentration exponentially increases with temperature (the Arrhenius stance[3-11]); b) a vacancy-free crystallite, regardless of size and shape, is thermodynamically stable only at 0 K[9,11,12]; and c) vacancy concentration is higher[13-17](lower[17,18]) in nanocrystals as the free energy of vacancy formation decreases[13-17](increases[17,18]) due to the size effect[13-23]. These findings have been successfully used in fields such as atomic diffusion[24,25], ionics[26,27], and electronics[28,29] for modeling and quantitative analysis of the many solid state physicochemical processes taking place in sizable single crystals.

However, even after careful compensation for the size effects[13-23], the statistical predictions related to small crystallites have yet to be verified by the experimental results or through ab initio models[13-16,23]. Point defects are nucleation centers for major crystal defects[30-32], and valid statistical information on them, particularly on vacancies, is essential for describing morphologies in micro- and nano-crystallites grown in different conditions. Mainstream defect statistics have not offered solid clues to these basic problems.

In accordance to previous reports, the present investigation begins with the commonly employed statistical thermodynamic relationship for the formation free energy of vacancies[5-7,9,10], but we prove that using Stirling's approximation[33-36] for simplifying the related mathematical equations leads to erroneous results in the case of nanocrystals. Then, we use the "improved Stirling's approximation" to establish a novel statistical model for crystal point defects. The obtained results are significantly different from those previously articulated. The results are valid for all point defects, but this work focuses on vacancies because of the relative abundance of experimental data on their formation free energy in a number of elemental crystals.

## 2. Results and Discussion

### 2.1 Vacancy Statistics: a Brief Critique

An atomic crystallite formed with $n_0$ atoms contains $n_v$ vacancies at thermal equilibrium at T K. The formation energy of a vacancy is $E_v$ and the change in the Gibbs free energy due to the formation of $n_v$ vacancies is given by[5-7,9,10]

$$\Delta G = n_v E_v - T\left(\Delta S_{conf} + n_v \Delta S_{vib}\right) \tag{1}$$

wherein $\Delta S_{conf}$ is the configurational entropy of $n_v$ vacancies in a crystallite of $n_0 + n_v$ lattice points, and $n_v \Delta S_{vib}$ is the change in the vibrational entropy of the $n_0$ atoms due to the formation of $n_v$ vacancies. $\Delta S_{conf}$ is calculated based on the fundamental definitions in the statistical thermodynamics[37,38]:

$$\Delta S_{conf} = k \ln \frac{(n_0 + n_v)!}{n_0! n_v!} \tag{2}$$

$$= k\left[\ln(n_0 + n_v)! - \ln n_0! - \ln n_v!\right] \tag{3}$$

where $k$ is the Boltzmann constant. Replacing equation (3) in equation (1) for $\Delta S_{conf}$ results in

$$\Delta G = n_v\left(E_v - T\Delta S_{vib}\right) - kT\left[\ln(n_0 + n_v)! - \ln n_0! - \ln n_v!\right] \tag{4}$$

Equation (4) has been the starting point of all analytical calculations of vacancy and other point defects concentrations in crystals. These calculations are continued by replacing the simple Stirling's approximate expression[33-36] ($\ln N! \approx N \ln N - N$) for all the logarithmic terms in equation (4). This approximation results in a size-independent expression for the normalized $\Delta G$

$$\frac{\Delta G}{n_0 kT} = \theta c_v + c_v \ln c_v - (1 + c_v) \ln (1 + c_v) \tag{5}$$

in which

$$c_v = \frac{n_v}{n_0} \tag{6}$$

and

$$\theta = \frac{E_v}{kT} - \frac{\Delta S_{vib}}{k} = \frac{G_f}{kT} \tag{7}$$

where $G_f$ is the formation free energy of a vacancy in the crystal. Based on equation (7), quantitative conditions imposed on $\theta$ can be translated into restrictions on the equilibrium temperature. At thermodynamic equilibrium, $n_v$ minimizes the normalized free energy of the crystallite. Hence, by taking the first derivative of the right hand side of equation (5) with respect to $c_v$ and equating it to zero the vacancy concentration at thermodynamic equilibrium, $c_{v,eq}$, is obtained[6,7,11,12,39]:

$$c_{v,eq} = \frac{1}{e^\theta - 1} \approx e^{-\theta} \tag{8}$$

According to equation (8), the obtained vacancy concentration is independent from $n_0$ (crystallite size) and has an Arrhenius-like temperature dependence. This is the relationship commonly utilized for point defect calculations in crystals[5-10,12,39].

However, using simple Stirling's approximation for $\ln N!$, when $N < 50$, may lead to significant errors[33-36,40]. Inserting the numerical values of $\theta$ related to real crystals (see below) in equation (8) results in $n_v / n_0$ values of at most $10^{-5}$ at feasible working temperatures. As a result, for nanosized crystallites with $n_0$ less than $5 \times 10^6$, the predicted vacancy number is less than 50. Hence, using simple Stirling's approximation for $\ln n_v!$ in equation (4) can lead to erroneous conclusions on nanocrystals. Moreover, concluding $n_v = 0$ at $T = 0\,K$, from equation (8), which is common in the literature, is erroneous as the approximations leading to equation (8) lose their validity for $n_v < 50$.

## 2.2 Novel Approach

Like the reviewed reports, our analysis starts from equation (4), wherein simple Stirling's approximation is utilized to replace $\ln(n_0 + n_v)!$ and $\ln n_0!$ terms. However, the "improved Stirling's approximation"[33-36,40], i.e.

$$\ln N! \approx \ln \sqrt{2\pi} + \frac{1}{2}\ln N + N\ln N - N \qquad (9)$$

replaces the $\ln n_v!$ term. The errors of utilizing equation (9) are in the acceptable range[40] for $N > 2$. (The even more accurate Gosper's approximation[33,41] was also tested, but analytically solving the equations involved proved difficult. Numerical solutions lead to results similar to those presented below.) The resulting relationship is different from equation (5) in nature:

$$\frac{\Delta G}{n_0 kT} = \frac{1}{2n_0} \ln(2\pi n_0 c_v) + \theta c_v + c_v \ln c_v - (1+c_v)\ln(1+c_v) \qquad (10)$$

wherein all parameters have the same significance as in equation (5). The equilibrium vacancy concentration would minimize $\Delta G$ and, hence, $c_{v,eq}$ is obtained from solving

$$\frac{\partial \Delta G}{\partial c_v} = 0 \qquad (11)$$

Equation (11) reduces to

$$\theta + \frac{\mu}{c_v} - \ln\frac{1+c_v}{c_v} = 0 \qquad (12)$$

in which

$$\mu = \frac{1}{2n_0} \qquad (13)$$

Equation (12) has a meaningful solution only if the following condition is satisfied,

$$\theta < \mu - 1 - \ln\mu \qquad (14)$$

which sets a conditional relationship between the equilibrium temperature and crystallite size. If inequality (14) is not satisfied the free energy of the crystallite, given by equation (10), would continuously increase with $c_v$, i.e. the lowest free energy is achieved by the vacancy-free crystallite. On the other hand, in conditions fulfilling inequality (14), equation (12) would render an exact solution in the following form:

$$c_{v,eq} = -\frac{\mu}{\mu + W_0\left(-\mu e^{\theta-\mu}\right)} \tag{15}$$

in which, $W_0$ is the main branch of the Lambert function[42-44]. The vacancy concentration given by equation (15) minimizes the free energy of the crystallite and, thus, determines its vacancy concentration at the thermodynamic equilibrium. At a constant temperature, equation (15) predicts lower defect concentrations for smaller crystallites (see below). Significant features of the condition (14) and the solution (15) are listed below.

**A)** For large crystallites, equation (15) reduces to the relationship commonly utilized for the point defect concentration calculations in the background literature[6,7,11,12,39]

$$\lim_{n_0 \to \infty} c_{v,eq} = \frac{1}{e^\theta - 1} \approx e^{-\theta} \tag{16}$$

Thus, the presented statistics approve Arrhenius-like temperature dependence only if: a) condition (14) is fulfilled and b) the crystallite under investigation is large enough to allow the approximate reduction of equation (15) to equation (16); the results of equation (15) obtained for $n_0 = 10^{10}$ still significantly differ from those of equation (8). Hence, equation (15) provides more accurate defect statistics than the commonly used equation (16) for micro-crystallites.

**B)** The complement of condition (14) leads to a conditional relationship between the crystallite size ($n_0$) and its temperature in the form of

$$n_0 < n^* \tag{17}$$

where $n^*$ is a dimensionless function of the equilibrium temperature:

$$n^* = -\frac{1}{2W_0\left(-e^{-\theta-1}\right)} \tag{18}$$

According to condition (17), at thermodynamic equilibrium, crystallites with a smaller number of atoms than $n^*$ are vacancy-free while larger ones contain defect concentrations given by equation (15). It is concluded that a crystal nucleus growing from vapor or solution, at T, would grow vacancy-free up to a size of $n^*$. Beyond this critical size, vacancy formation becomes thermodynamically favorable and, at this stage, equation (15) determines the equilibrium defect concentration. Based on the available $E_v$ and $\Delta S_{vib}$ data for different crystals[45,46], the $n^*$ values of a number of elemental crystals including Al, Au, Cu, Si and Ge are calculated and plotted vs. the equilibrium temperature in Fig. 1a which depicts larger $n^*$ values at lower equilibrium temperatures. This implies that larger defect-free nanocrystals can be grown at lower temperatures. For instance, according to Fig. 1a, gold nanocrystals can grow defect-free up to $n_0 = 4\times10^4$ close to thermodynamic equilibrium at 1000 K, while at 300 K defect-free growth of gold crystallites containing $n_0 = 6\times10^{15}$ appears feasible. For the crystallites larger than these critical sizes, the equilibrium vacancy concentration is predicted using equation (15).

**C**) It is customary, though mathematically incorrect, to conclude from equation (8) that the equilibrium vacancy concentration would reduce to zero only if the temperature approaches 0 K[9,11,12]. Variations of $c_{v,eq}$ with respect to the equilibrium temperature for nanocrystals of Al,

Au, Cu and Si, each at three different sizes, are shown in Fig. 1b. According to these diagrams, a vacancy-free nanocrystal is at thermodynamic equilibrium up to elevated temperatures depending on its size. Smaller crystallites can be vacancy-free and at thermodynamic equilibrium up to higher temperatures. For instance, a vacancy-free gold crystallite of $n_0 = 10^8$ is at thermodynamic equilibrium up to 585 K. For a vacancy-free silicon nanocrystal of the same size the equilibrium temperature range is up to 1288 K (Fig. 1b).

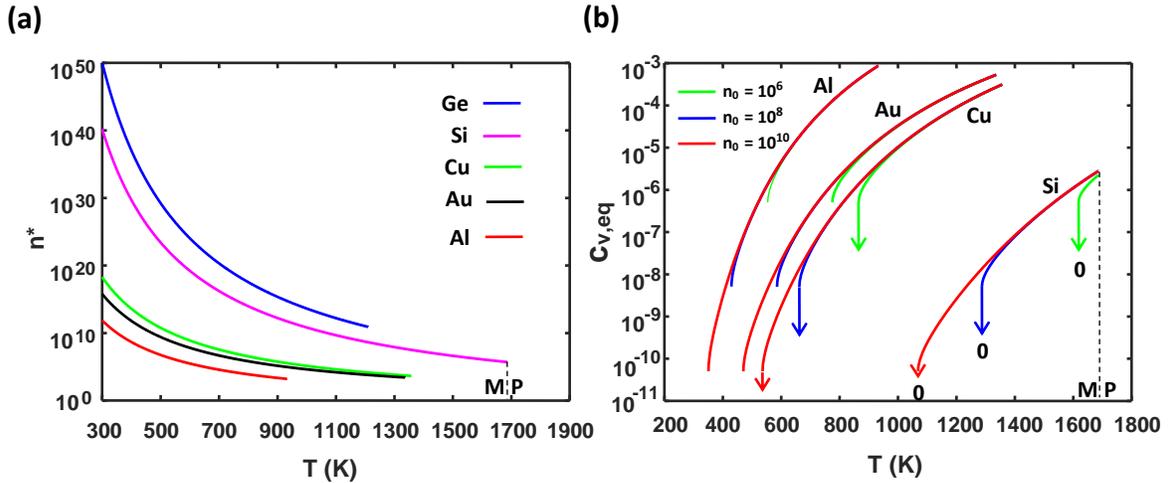

**Figure 1.** Illustration of the results of the presented vacancy statistics; a) variations of the largest vacancy-free crystallite of the stated materials at thermodynamic equilibrium ($n^*$) vs. the equilibrium temperature from 300 K to the materials' melting points, and b) variations of the equilibrium vacancy concentration ($c_{v,eq}$) in the crystallites of the stated materials and sizes with respect to the equilibrium temperature.

**D)** In the vacancy population calculations via equation (15) for real crystals, we have applied the available bulk $E_v$ data to the whole volume of a crystallite. The fact that vacancy formation energy decreases from the bulk to the surface may raise questions in the validity of the results of equation (15) when applied to smaller crystallites. However, the climax of our predictions occurs for a crystallite of $n^*$ atoms; below this size the predicted vacancy concentration is zero. As shown in Fig. 1a, $n^*$ at room temperature for all materials examined is larger than $10^{10}$. The fraction of the surface atoms in a crystallite of $n_0$ atoms is approximately given by[47,48]

$$F \approx \frac{4}{\sqrt[3]{n_0}} \tag{19}$$

which results in 0.002 for a crystallite of $n^*$ atoms. Considering that only the $E_v$ values of the two layers beneath the surface are significantly affected by the surface effect[49,50], the $E_v$ adjustment applies, at most, to 0.006 fraction of the lattice sites. The modification required is minor and affects only the numerical value of $n^*$; the concept of having zero vacancy in crystallites with $n_0 < n^*$ remains intact.

## 2.3 Vacancy Statistics in Nanocrystals

The equilibrium concentrations of vacancies and other point defects in nanocrystals have generally been calculated based on equation (8) by considering the size effect-related corrections on the bulk $E_v$ and $\Delta S_{vib}$ values[13]. Assuming both parameters to decrease with the particle size, these calculations predict considerably higher vacancy concentrations in nanosized particles than the bulk material[13]. However, using condition (14) and equation (15) instead of equation (8) leads to very different results.

The $n^*$ values calculated for a number of atomic crystals including Al, Au, and Cu fall within the nanometric size range (see Fig. 1a). Accordingly, $E_v$ and $\Delta S_{vib}$ have to be corrected for the size effect before insertion in condition (14) and equation (15). For a spherical crystallite with diameter $D$, the commonly utilized approximate relations are[13,14]

$$E_v(D) = E_v\left(1 - \frac{\alpha_{shape}}{D}\right) \qquad (20)$$

$$\Delta S_{vib}(D) = \Delta S_{vib}\left(1 - \frac{\alpha_{shape}}{D}\right) \qquad (21)$$

wherein $\alpha_{shape}$ is a constant of the crystallite geometry[13,51-53]. The suggested $\alpha_{shape}$ for the spherical nanoparticles of the selected materials is around 1.6 nm[13,53]. These approximate relations are applicable to particles with more than 3000 atoms[14]; by utilizing these relations we are assuming $n_0$ to be larger than 3000. For such a nanocrystal, according to the "liquid drop model"[54], we have

$$D = \left(\frac{6}{m\pi}\right)^{\frac{1}{3}} a_0 \, n_0^{\frac{1}{3}}$$

(22)

in which, $a_0$ is the lattice constant of the crystal and m is the number of atoms per unit cell. Replacing equation (22) for D in equation (20) and equation (21) results in the size effect-compensated $E_v$ and $\Delta S_{vib}$. Their insertion in condition (14) elucidates the condition for the free energy of defect formation to have a minimum:

$$n_0 > n^{**} \tag{23}$$

in which

$$n^{**} = -\left[\frac{\alpha_{shape}}{3\gamma a_0} \cdot \frac{\theta}{W_0\left(-\frac{\alpha_{shape}}{3\gamma a_0}\sqrt[3]{\frac{2}{e}}\,\theta e^{-\frac{\theta}{3}}\right)}\right]^3 \tag{24}$$

and

$$\gamma = \left(\frac{6}{m\pi}\right)^{\frac{1}{3}} \tag{25}$$

The $n^{**}$ as determined via equation (24) is the size effect-modified $n^*$, i.e. for particles with $n_0 < n^{**}$ atoms, the free energy of vacancy formation has no minimum. It is concluded that a nanocrystal growing close to thermodynamic equilibrium will grow vacancy-free up to the critical size of $n^{**}$. The variations of the $n^{**}$ of Al, Au, and Cu with respect to the equilibrium temperature are shown in Fig. 2; those related to Si and Ge turned out to be close to their respective $n^*$ values shown in Fig. 1a.

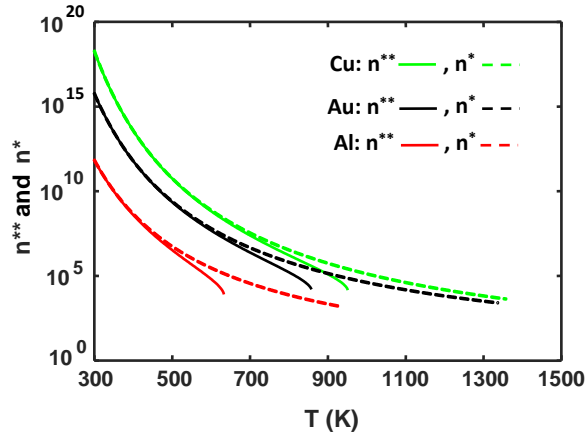

**Figure 2.** The variations of $n^{**}$ with respect to the equilibrium temperature for the stated materials; $n^*$ is provided for comparison.

In a crystallite larger than $n^{**}$, the $c_{v,eq}$ is calculated by inserting its size and size effect-corrected $\theta$ in equation (15). In stark contrast with the former predictions, the results depict lower equilibrium defect concentrations in smaller crystallites. The variations of $c_{v,eq}$ with respect to particle size obtained for spherical particles of Au, Cu, Pt, Al, Si and Ge each at two different temperatures, are presented in Fig. 3a-l, where they are compared with the similar predictions based on the Arrhenius stance.

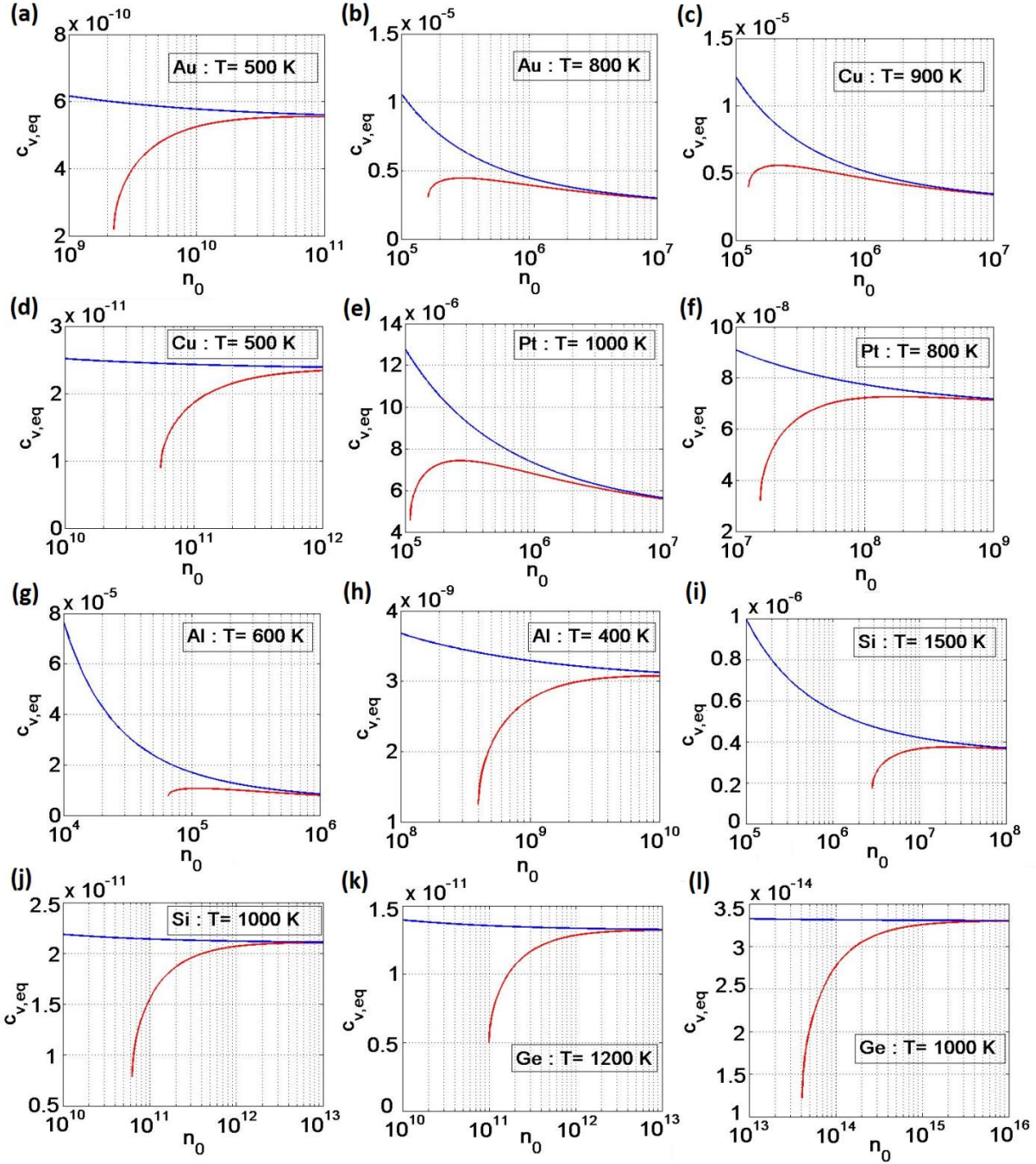

**Figure 3.** Predicted variations of vacancy concentration with crystallite size (red) is compared with those calculated based on the Arrhenius stance (blue) for Au (a and b), Cu (c and d), Pt (e and f), Al (g and h), Si (i and j), and Ge (k and l) at the stated equilibrium temperatures.

In all examples produced in Fig. 3, our novel statistics predict a completely different trend in the defect population variations for the nanocrystals, but its results approach those of the mainstream defect statistics when applied to larger crystallites. The predicted trend of variations depends strongly on the equilibrium temperature as clarified by the different temperature frames related to the same material in Fig. 3. For instance, for Si at 1000 $K$, the novel statistics predict a considerable drop in the vacancy population for crystallites with smaller than $10^{12}$ atoms in comparison to the vacancy population in bulk (Fig. 3j). The same drop occurs in crystallites with smaller than $10^7$ atoms when the equilibrium temperature is raised to 1500 $K$ (Fig. 3i). The intricate interconnections between the equilibrium defect concentration, equilibrium temperature, and crystallite size are depicted in the 3-D presentations provided for Al, Cu, Au, and Si in Fig. 4a-d.

Our mathematical work, predicting drastically smaller vacancy concentration in nanocrystals than bulk, describes the results of the plastic deformation measurements[55] carried out on the nanometer-sized crystals of Au, Pt, W, and Mo, for the first time. Indeed, the results of this ingenious experimental work[55] verify our theoretical predictions.

**(a)**

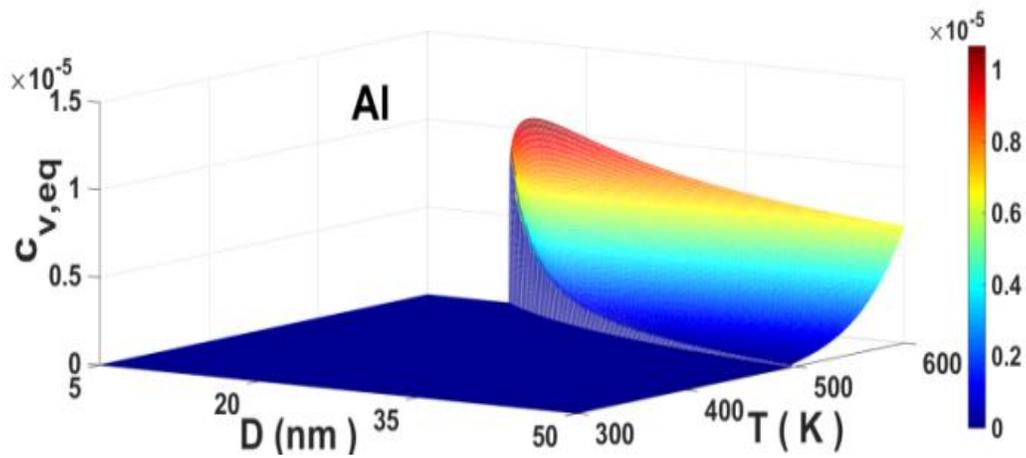

**(b)**

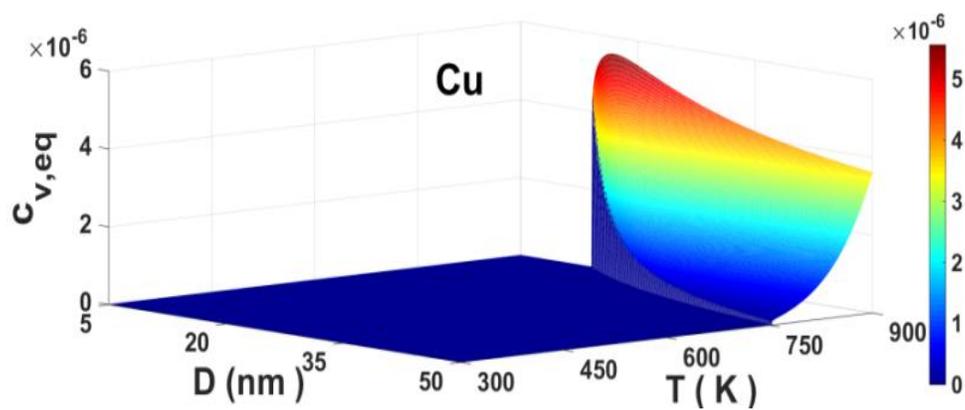

**(c)**

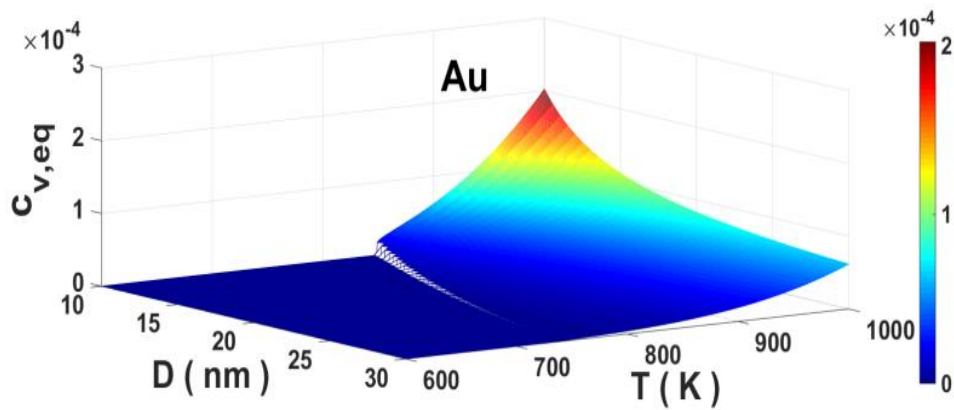

**(d)**

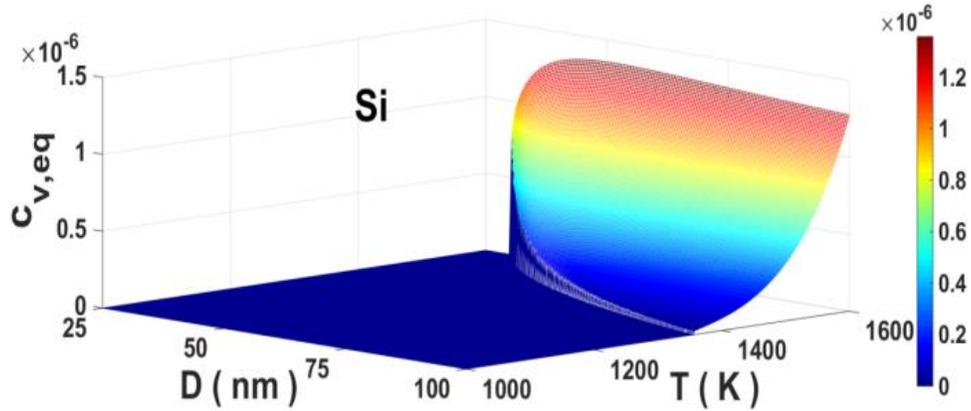

**Figure 4.** Variations of the equilibrium vacancy concentration with respect to the crystallite diameter and equilibrium temperature for Al (a), Cu (b), Au (c), and Si (d). In each frame, the dark blue plateau illustrates the size and equilibrium temperature ranges for the growth of defect free crystallites.

## 3. Conclusions

We established an exact relationship between the equilibrium vacancy concentration in a crystallite and a Lambertian function of its size and temperature. This relationship can also be utilized for determining the equilibrium concentrations of other point defects based on their respective formation free energies. According to this relationship, and contrary to the currently accepted thermodynamic model, a vacancy-free crystallite can be at thermodynamic equilibrium at temperatures well above 0 $K$. For example, the equilibrium vacancy concentration in a silicon nanocrystal with $10^5$ atoms was shown to be zero up to its melting point. The Arrhenius-like temperature dependence of defect concentration was clarified to be valid only for large crystals.

These findings are anticipated to have profound implications in our understanding of the morphological occurrences in crystallites grown in different conditions. A crystallite, growing from a vapor or solution at around thermodynamic equilibrium well below its melting point, is predicted to grow defect-free up to a critical size. Further growth would involve point defect generation, which can initiate the nucleation of crystallites with different orientations, which, in turn, would grow defect-free up to the critical size. Similar arguments can be made for impurity diffusion and surface poisoning which would stop growth around the critical size resulting narrow particle size distributions. Critical sizes can be determined for different materials and growth conditions by inserting the experimental data available in the literature regarding the formation free energy of the specified defect into the relationship developed here; practical examples were provided. For all the materials examined, the equilibrium vacancy concentration is less in smaller crystallites. This trend holds regardless of the positive or negative size-related free energy of formation adjustments in nanocrystals. Our results show that nanocrystals are mostly defect-free; selecting lower growth temperatures ensures perfection.

**Authors' contributions**

FHB defined the project, supervised its progress, and wrote the manuscript; HRSS carried out the mathematical work and literature survey and produced the figures. The idea of using two different Stirling's approximation in the main statistical relationship occurred in a discussion between the authors.

**Competing financial interests:** The authors declare no competing financial interests.